\documentclass[preprint,12pt]{elsarticle}

\usepackage{amsmath,amssymb,amsfonts}
\usepackage{algorithmic}
\usepackage{graphicx}
\usepackage{textcomp}
\usepackage{xcolor}
\usepackage{enumitem}
\usepackage{url}
\usepackage{booktabs}
\usepackage{color}
\usepackage{colortbl}
\usepackage{capt-of}
\usepackage{tabu}
\usepackage{scalerel}
\usepackage{tikz}
\usepackage{comment}
\usetikzlibrary{svg.path}
\usepackage{enumitem}
\usepackage{amsthm}

\usepackage[T1]{fontenc}

\usepackage{listings}

\usepackage{float}
\usepackage{hyperref}

\newtheoremstyle{mystyle}
  {}
  {}
  {\itshape}
  {}
  {\bfseries}
  {.}
  { }
  {\thmname{#1}\thmnumber{ #2}\thmnote{ (#3)}}

\theoremstyle{mystyle}

\newtheorem{Req}{Requirement}

\usepackage{comment}

\journal{Arxiv}

\begin{document}

\begin{frontmatter}



\title{Dynamic and Scalable Data Preparation for Object-Centric Process Mining}


\author[inst1]{Lien Bosmans}

\affiliation[inst1]{organization={Business \& Data Analyst, Randstad Digital},
            city={Leuven},
            country={Belgium}}

\author[inst2]{Jari Peeperkorn}
\author[inst2]{Alexandre Goossens}
\author[inst3]{Giovanni Lugaresi}
\author[inst2]{Johannes De Smedt}
\author[inst2]{Jochen De Weerdt}

\affiliation[inst2]{organization={Research Center for Information Systems Engineering (LIRIS), KU Leuven},
            city={Leuven},
            country={Belgium}}

\affiliation[inst3]{organization={Centre for Industrial Management / Traffic and Infrastructure (CIB), KU Leuven},
            city={Leuven},
            country={Belgium}}

\begin{abstract}
Object-centric process mining is emerging as a promising paradigm across diverse industries, drawing substantial academic attention. To support its data requirements, existing object-centric data formats primarily facilitate the exchange of static event logs between data owners, researchers, and analysts, rather than serving as a robust foundational data model for continuous data ingestion and transformation pipelines for subsequent storage and analysis. This focus results into suboptimal design choices in terms of flexibility, scalability, and maintainability. For example, it is difficult for current object-centric event log formats to deal with novel object types or new attributes in case of streaming data. This paper proposes a database format designed for an intermediate data storage hub, which segregates process mining applications from their data sources using a hub-and-spoke architecture. It delineates essential requirements for robust object-centric event log storage from a data engineering perspective and introduces a novel relational schema tailored to these requirements. To validate the efficacy of the proposed database format, an end-to-end solution is implemented using a lightweight, open-source data stack. Our implementation includes data extractors for various object-centric event log formats, automated data quality assessments, and intuitive process data visualization capabilities. 
\end{abstract}




\end{frontmatter}


\section{Introduction}\label{sec:introduction}

Introducing the notion of objects in process mining relaxes the single object (case) assumption, thereby enabling the description, analysis, and monitoring of complex, interdependent processes. This shift provides clear opportunities for scenarios where multiple viewpoints need to be considered simultaneously. For instance, in production companies, integrating both material and equipment viewpoints and cardinality relationships between components and assembled parts is crucial for effective planning and scheduling. Recent advancements have introduced various formats for object-centric event data, each offering design choices and relational schemas to facilitate the storage of a wide variety of process data. However, many existing formats overlook the critical perspective of data sourcing and preparation, in particular in dynamic environments. Introducing new object types, event types, modifying their attributes, or even updating their meta-information, can lead to disruptive changes in data pipelines that rely on rigid table and column structures. Alterations such as table deletions, column renaming, or even the addition of new columns can cause data ingestion issues, potentially necessitating full table refreshes before incremental data updates can resume, a time-consuming process prone to manual intervention. These challenges highlight the need for robust database implementations capable of supporting dynamic environments with continuous process monitoring and improvement, a gap that currently limits broader adoption of object-centric process mining in industry. \\

\begin{figure}[!htbp]
    \centering
    \includegraphics[width=0.90\textwidth]{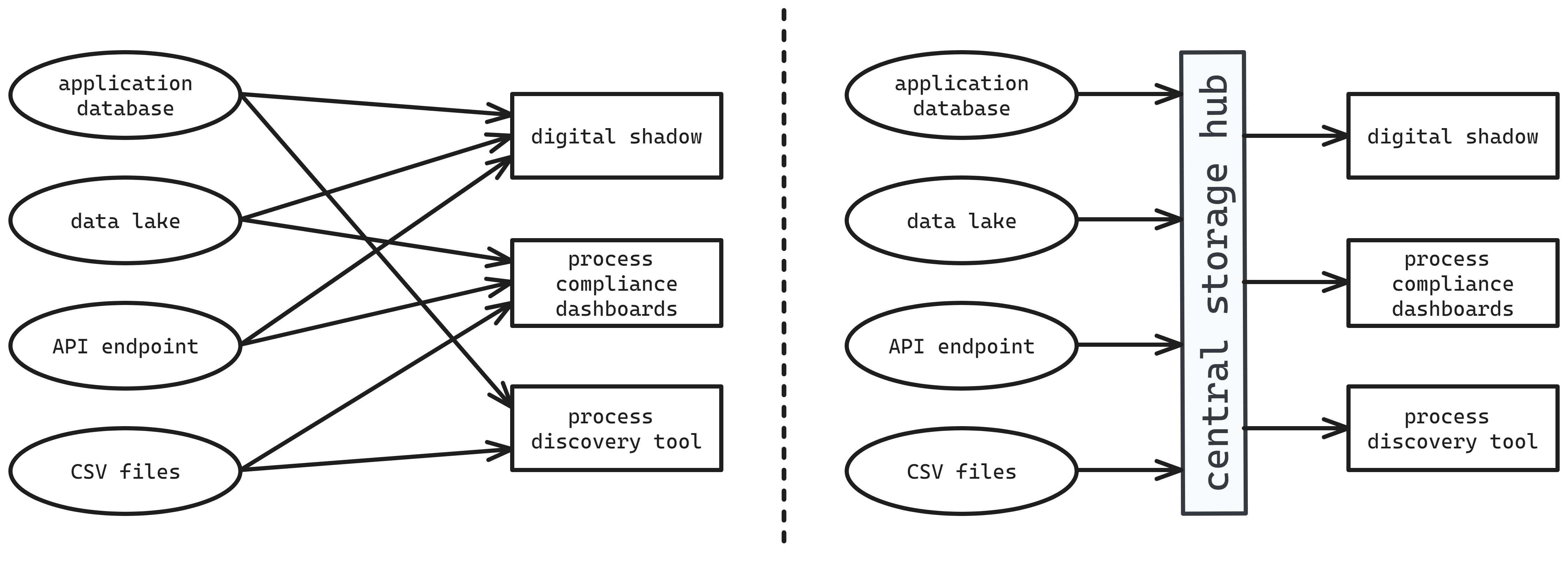}
    \caption{Example of point-to-point (left) vs. hub-and-spoke (right) architecture. In a point-to-point architecture, every application (rectangle) is directly connected to the required data sources (oval). A hub-and-spoke architecture introduces an abstraction layer (hub) that separates data sources from their use cases. Only one connection (spoke) needs to be updated when changes are made to a data source or application.}
    \label{fig:hub_spoke}
\end{figure}

This paper proposes a \emph{hub-and-spoke} architecture to maintain a flexible and scalable database format for object-centric event data.
In this architecture, data sources connect to a central hub rather than directly to applications, significantly reducing system complexity, as shown in  Fig.~\ref{fig:hub_spoke}.
This model allows organizations to establish data ingestion processes once, with the flexibility to update connections to individual data sources or applications as needed.
Additionally, this work proposes a relational schema that serves as a database format for object-centric event data. Current data formats typically facilitate static event log exchange between data owners, researchers, and analysts, rather than functioning as the foundational data model for continuous data ingestion, storage, and transformation pipelines. Although existing object-centric data formats like OCEL 2.0~\cite{berti2024ocel} effectively conceptualize object-centric event logs and  serve as an exchange format for OCPM tools, they often lack direct implementation capabilities. To address these limitations, our schema design prioritizes practical deployment, flexibility, maintainability, and scalability. It forms the cornerstone of our hub-and-spoke architecture for object-centric event data. Importantly, our data storage hub architecture remains fully compatible with state-of-the-art data formats such as OCEL 2.0, facilitating seamless transformation between different data formats via importers and extractors. \\

In summary, this work contributes to existing literature by:
\begin{itemize} 
    \item Establishing a comprehensive set of requirements for robust object-centric event log storage, data ingestion, and preparation in dynamic environments, emphasizing functionality, scalability, maintainability, and generalizability. 
    \item Introducing a relational schema that serves as the cornerstone of the hub-and-spoke architecture for object-centric event log storage. This schema is designed to seamlessly accommodate continuous changes in process data architecture, including novel object types, event types, or attributes, while ensuring compatibility with leading formats such as OCEL 2.0.
    \item Demonstrating the practical implementation of the relational schema through an end-to-end solution using a lightweight, open-source data stack. This implementation is capable of effectively ingesting process data from divers sources. 
    \item Extending the implementation with data extractors tailored for different object-centric event log formats, automated data quality tests, and visualizations resembling event knowledge graphs.
\end{itemize}

To that end, this work contributes to the existing literature by focusing more on dynamically ingesting event data from various data sources, storing the event data in a robust manner, and allowing the on-demand generation of object-centric event logs in various formats. \\

The remainder of the paper is organized as follows:
Section~\ref{sec:related_work} reviews related work on object-centric event log formats and data preprocessing pipelines. Section~\ref{sec:problemstatement} outlines requirements motivating our flexible and scalable database format, while section~\ref{sec:relational_schema} details our schema design choices.
Section~\ref{sec:tool} introduces our implementation, and section~\ref{sec:conclusion} concludes this work, while suggesting avenues for future research.


\section{Related Work}\label{sec:related_work}
In this section an overview of the current object-centric event storage formats is provided, followed by relevant literature related to hub-and-spoke and data pipeline architectures.

\subsection{Object-Centric Event Log Formats}
Object-centric process mining has gained increasing attention in recent years~\cite{berti2023advancements}. The concept of representing multiple objects in processes was first proposed as early as 2001 with proclets~\cite{van2001proclets}.
However, formalization of object-centric event log formats emerged much later, beginning with the introduction of the eXtensible Object-Centric (XOC) logs \cite{van2017object} in 2017 and the accompanying  Object-Centric Behavioral Constraint model (OCBC)~ \cite{li2017automatic}.
Unfortunately, XOC logs face scalability issues due to  storing the complete object model with each event that updates an object or the object model. In response, the Object-Centric Event Log (OCEL) 1.0 format was proposed, which separated objects and events into separate tables, linking objects directly to the events they participate in~\cite{ghahfarokhi2020ocel}. 
Despite improvements in scalability, OCEL 1.0 lacked support for traceable and flexible object attribute changes, requiring the creation of new object versions with each update without version tracking. Moreover, it does not store object relations and thus cannot handle changing object relations over time.\\

Subsequent advancements addressed these shortcomings. The Data-aware Object-Centric Event Log (DOCEL) format addressed the traceability of object attribute changes by introducing foreign key links between attribute values, events, and objects to track attribute changes effectively~\cite{goossens2022enhancing}.
Similarly, the Artefact-Centric Event Log (ACEL) \cite{m2023process} format accommodates dynamic attribute and object relation changes, although its departure from relational normalization principles complicates deeper analytical insights~\cite{10.7551/mitpress/12274.003.0034}.
More recently, the same task force responsible for the eXtensible Event Stream (XES) format~\cite{verbeek2010xes} introduced the Object-Centric Event Data model (OCED) as a meta-model for storing object-centric process data, yet lacks a formalized event log format\footnote{https://www.tf-pm.org/resources/oced-standard}.
However, the Object-Centric Event Log 2.0 (OCEL 2.0) standard has aligned with OCED principles by incorporating  attribute value timestamps and supporting static object relations. \\

The previous formats, however, mainly serve as an exchange format or can only be used for real-time logging in a context where the process data structure does not change along the underlying system life cycle, i.e., the tables and columns for object types, event types and attributes need to be determined upfront. This is known to cause maintainability and scalability challenges in industry, which has also been discussed in literature concerning data interoperability and database management~\cite{leah2024maximizing}. \\

Another innovative approach is the use of Event Knowledge Graphs (EKG)~\cite{fahland2022process}, which store object-centric event data as flexible knowledge graphs. Since the underlying data storage is a graph database, therefore lacking a schema implementation, this approach is adaptable and allows for extending with new types and concepts as needed. EKG's are used as a data store (hub) in PromG, a Python library for interactively querying, exploring and visualizing object-centric event data \cite{swevels2023object}. While the implementations of EKG extensively consider \textit{using} the stored data, the formulated requirements in \cite{esser2021multi} and subsequent work fail to take into account the engineering challenges of the data \textit{preparation} stage. This is not limited to data ingestion; for non-graph applications it is also difficult to query the data in an EKG data store directly, necessitating additional data wrangling.

\subsection{Hub-and-Spoke and Data Pipelining Architectures}

The use of hub-and-spoke in data architectures has a long tradition that can be traced back to Inmon introducing the data warehouse in 1989 \cite{reis2022fundamentals}.
Already before the existence of modern data tools, the importance of a correct and scalable data pipelining approach was recognized in \cite{simmhan2009building}.
The term \textit{pipeline jungle} was introduced to describe a type of technical debt often found in the data preparation stage for ML applications~\cite{sculley2015hidden,obrien202223} and advocates the need for an integrated approach to manage data preprocessing.
This \textit{pipeline jungle} was also reported in \cite{raasveldt2020data} with many data analysis approaches gluing multiple data tools together with scripting languages because relational database management systems (RDBMS) do not provide an integrated solution for their needs, and propose an embedded analytics system to fill this gap.
Additionally, \cite{shivashankar2022maintainability} report that despite data preprocessing being a highly experimental step it does require reproducibility once implemented, which is often not guaranteed due to correction cascade. Therefore, one solution is to implement better tooling and infrastructure to maintain and update models in an iterative fashion.  \\

Recent work provides a high-level overview of data pipeline architectures and identifies multiple factors that can influence their ability to provide quality data, including dependencies between processing steps and a (lack of) modularization of data flow architectures~\cite{foidl2024data}.
This is supported by \cite{reis2022fundamentals}, who suggest \textit{Build loosely coupled systems} as a key principle for good data architecture. Moreover, it explains how even minor schema updates can lead to unacceptable delays in projects as they might impact the whole data pipeline.

\section{Problem Statement}\label{sec:problemstatement}

Preparing data for analytical purposes in an industry context is inherently dynamic. The focus often lies on near real-time data availability, as opposed to analyzing historical data sets. Changing project scopes and agile methodology regularly require integrating with new data sources containing novel processes, event and object types, or attributes. Finally, it can be difficult to keep up with new tools and technologies in a rapidly evolving data landscape - not unlike the active research field of object-centric process mining. This results in a balancing act between choosing the best tools available today, and avoiding vendor lock-in to keep future options open.  \\

As illustrated in Section \ref{sec:related_work}, we believe that for this dynamic context, the structures provided by current OCPM data formats are too rigid to optimally capture the flexibility that is required for changing database schemes, organic evolution and proliferation of data sources, and the seamless integration of new object types, event types, and features.
In response to these challenges, it is imperative to define comprehensive requirements for object-centric event data storage and analysis. These requirements should emphasize flexibility, scalability, and agility in data ingestion, preparation, and storage to support object-centric process mining applications in realistic yet inherently dynamic environments. Therefore, these requirements are organized into three key dimensions: (1)~maintainability and scalability, (2)~flexibility and functionality, and (3)~generalizability.

\subsection{Maintainability and Scalability}
Data preparation and system integration pose significant challenges for organizations adopting process mining \cite{Martin2021}.
While data acquisition, data preparation, and event log storage are recognized as distinct stages of preprocessing, iterative feedback loops often necessitate revisiting each phase~\cite{biswas2022art}.
Agile organizations, for instance, typically begin with a limited scope, adding more event types, object types, attributes and relations over time. Determining the granularity for event and object types upfront can be challenging. Initial insights may prompt organizations to refine or merge certain types, such as consolidating `cash payment' and `card payment' into `payment', or using `car' and `truck' instead of `vehicle'. 
Such changes may require reorganizing existing attributes and relations. Even minor modifications such as renaming a type or attribute, though seemingly insignificant to end users, can cascade into code-level impacts if not anticipated.
Real-time system integration introduces further complexities, including data integration and ingestion challenges and compatibility issues \cite{foidl2024data}.
Adopting a hub-and-spoke architecture reduces system interactions within an organization \cite{foidl2024data}.
However, to effectively address these challenges, the central storage hub must support continuous data ingestion through batch processing or streaming.
Even in established processes, unforeseen object or event types, or attributes can arise. Ideally, the data model should accommodate these changes seamlessly without necessitating schema adjustments.
Otherwise, downstream data pipelines reliant on schema-defined pointers from the central hub require manual intervention to ensure proper execution. This leads to the identification of the first requirement. 

\begin{Req}[Req\ref{Req:Robust}] \label{Req:Robust} 
\textbf{Robust Data Model}
\begin{itemize} 
    \item The relational schema of the database should be agnostic of process-specific details such as names of event types, object types, attributes and relations.
    \item The data model should facilitate the continuous addition of new data without necessitating schema modifications
\end{itemize}
\end{Req}

Thus, all data should conform to the exact same tables and table columns, significantly enhancing the hub's maintainability. While this approach may appear rigid, it offers substantial flexibility by eliminating the need for re-engineering when introducing changes to the process data. No additional tables are required when adding new event or object types, and no columns need to be added, renamed, or deleted when changing the set of attributes linked to event or object types.  \\

Scalability encompasses several dimensions such as the capability to handle large datasets cost-effectively (whether on cloud or servers), elasticity to adjust systems according to organizational needs, and integration with maintainability considerations~\cite{alwadi2018toward}. The human capital cost associated with maintainability is crucial for scalability.
Compute costs constitute a significant part of cloud expenses. To mitigate compute costs, the relational schema should support efficient queries, such as minimizing joins between large tables~\cite{liu2023cost}.
Consideration of querying patterns is crucial during data model design~\cite{reis2022fundamentals}. For instance, the decision to store event timestamps in a separate table for every event type (activity), as in OCEL 2.0, necessitates expensive join operations with each of those tables to identify the previous or next event linked to a certain object.
Storage costs can be minimized using a space-efficient data model combined with modern storage formats that compress data effectively~\cite{reis2022fundamentals}.
To manage the growing volume of data captured by businesses, database partitioning enhances scalability and performance for large datasets. Ideally, related events and objects should be grouped within the same partition. Effective data partitioning dimensions depend on the process context; potential candidates include time, geographical location, or business unit, whereas event and object types are less suitable due to process flows typically involving a mix of these elements.
Data from different sources may become available at varying times;  while data ingestion synchronization across systems can be achieved with delays, a storage hub accommodating asynchronous data ingestion requires less engineering effort to set up, thereby enhancing scalability.
This is summarized in our second requirement.

\begin{Req}[Req\ref{Req:Scalable}] \label{Req:Scalable} \textbf{Scalable Data Storage}
    \begin{itemize} 
	 \item Accommodate process-related querying patterns to reduce expected compute cost of table joins.
     \item Support database partitioning to address scalability or performance issues.
     \item Ability to handle asynchronous data ingestion to simplify pipeline setup across diverse data sources, reducing dependencies.
 \end{itemize}
 \end{Req}

Graph databases, such as those used for EKGs, specialize in exploring relationships among various data items such as events and objects. However, their industrial adoption remains lower compared to traditional relational database management systems (RDBMSs), suggesting higher entry barriers \cite{vicknair2010comparison}. 
Recent work highlights that RDBMSs can outperform more specialized graph DBMSs for graph simulation up to 10$\times$ and predicts that RDBMSs will continue to provide and improve graph-centric APIs on top of SQL or via extensions \cite{stonebraker2024goes}.
Thus, a traditional relational schema remains a reliable and future-proof option for storing object-centric event data.

\subsection{Flexibility and Functionality}

In the data preparation stage, data moves from source systems, typically application databases that are highly flexible in nature, towards event logs suitable for process mining, removing ambiguity in the process. Since our proposed database sits between the source data and process mining applications, we argue that its format should primarily employ a more flexible relational schema than the strict assumptions needed for (object-centric) process mining algorithms. Indeed, process owners may need to extend functionalities beyond the requirements of a proposed object-centric data format to suit their specific needs.
This emphasis on flexibility can leave some ambiguity that needs to be resolved when extracting data for process mining. However, the robust data model ensures that it can be done rule-based and therefore automated.  \\

Below, a non-exhaustive list of additional functionalities is provided, that could be desirable in different scenarios, illustrated by simple examples.
These complement the requirements previously discussed in literature regarding data formats, such as allowing objects to be linked to multiple events and events to different objects \cite{goossens2024objectcentric,berti2023advancements}.
These additional requirements, often relaxations of more strict relationship types in OCED formats, may not be relevant for each process type but should be adaptable to comply with different formats if needed.
The first group of requirements pertains to the event-to-object relations:

\begin{Req}[Req\ref{Req:E2O}] \label{Req:E2O} \textbf{Event-to-Object Relationships} 
\begin{itemize} 
    \item Different types of relationships between an event-object pair can exist and should be distinguishable (direct and indirect).
    \item An event can be related to multiple object attribute value updates.
    \item The event timestamp and the timestamps of the object attribute changes can all differ.
    \item An update of an object attribute value can be related to multiple events.
\end{itemize}
\end{Req}

This introduces two types of relations. The direct event-to-object relationship indicates which objects are actively involved in an event and in what role. The indirect relation, via object attribute value updates, captures causal relationships between events and object attribute changes. Both are necessary in a common database format as they store different information about the process. The importance of each relation depends on context. For example, causal relations might be unidentifiable in Internet-of-things (IoT) data, severely limiting the usability of a format such as DOCEL, but are crucial in decision models. 
To illustrate the flexibility requirements for this relationship, consider banking applications. When executing a payment, the account balance of the receiving party is often updated hours after the event, indicating a need for non-matching timestamps. A single transaction event updates both the balances of sender and receiver. Conversely, money transfers below a certain amount do not warn investigation but multiple of those transactions close to the threshold can change a categorical account attribute to ``suspicious''. Thus, the relationship between events and object attribute changes can be many-to-many. 
Note that no assumption that it is feasible for organizations to add all process data instantaneously and automatically, is made. If this would be the case each transaction event can be matched to a certain count (object) attribute for the system to be able to know when to change the other categorical attribute to ``suspicious''. Each incremental update of the object attribute (count in this case) would then still be related to only one event. The categorical account attribute change to ``suspicious'' would in this case have to be linked to this count attribute change. To avoid this convolution,it was decided to allow an update of an object attribute value to be related to multiple events.  \\

For a second type of relationship, object-to-object, different relaxations and functionalities currently not allowed by all object-centric formats can also be desirable:

\begin{Req}[Req\ref{Req:O2O}] \label{Req:O2O} \textbf{Object-to-Object Relationships} 
\begin{itemize} 
    \item An object should be able to relate to multiple objects.
    \item Different types of relationships between objects can exist and should be distinguishable. 
    \item Object-to-object relationships can be temporal or dynamic, and these changes should be traceable over time.
\end{itemize}
\end{Req}

Consider an organizational chart where employees are modeled as objects. Managers have employees who report directly to them and may also report to another person. It is important to qualify these person-to-person relationships. However, organizational charts are not static: new employees start, others leave, and some move to different roles. When checking for irregularities, it is important to know which relationships existed at specific points in time. Most formats, except ACEL, do not allow object-to-object relationships to change over time, making it impossible to track changes accordingly. \textbf{}

Note that event-to-event relationships are not included here. This is supported by the observation that when an event can no longer be considered instantaneous and requires start/end or parent/child events, its data should be modeled as a temporal object instead.

\subsection{Generalizability}
The aim of using a standardized object-centric event log storage hub is to pull information from different systems, store it appropriately, and allow for the extraction of (filtered) parts of the event data. 
In line with maintainability and scalability, this hub would ideally require a one-time configuration and facilitate the extraction of event logs for analytical purposes as needed.
Considering flexibility and functionality, a database should be capable of storing all process-related information, even if most process mining applications do not require all of it (yet).
Ideally, the hub should support extractors for various formats, enabling process owners to share event logs in the most suitable format for the specific application and to use different algorithms designed for different specific formats.
Allowing the process owner to select the granularity level of the exported event logs can also be recommended. This might be of special interest when dealing with processes including e.g. IoT data. This leads to the following requirements:

\begin{Req}[Req\ref{Req:Generalization}] \label{Req:Generalization} \textbf{Universal Data Integration}
\begin{itemize} 
    \item An event log storage format should accommodate different sources and data types as input.
    \item Ideally, an event log storage format should generalize across various (object-centric) event log formats, offering flexible import and export extensions for different formats.
\end{itemize}
\end{Req}

While this last requirement might not always be strictly necessary, it can be convenient for several use cases to be able to extract different event log formats from such a central hub.

\section{Relational schema}\label{sec:relational_schema}
The previous section explored the design principles behind an intermediate storage hub tailored for object-centric event data.
In this section, a concrete implementation in the form of a relational schema definition is presented, providing a blueprint for incorporating object-centric event data into both new and existing data platforms. The relational schema supports append-only incremental data ingestion, ensuring new information can be seamlessly added without altering existing records in the database (in line with Requirement \ref{Req:Scalable}).

\begin{figure}[!ht]
    \centering
    \includegraphics[width=0.80\textwidth]{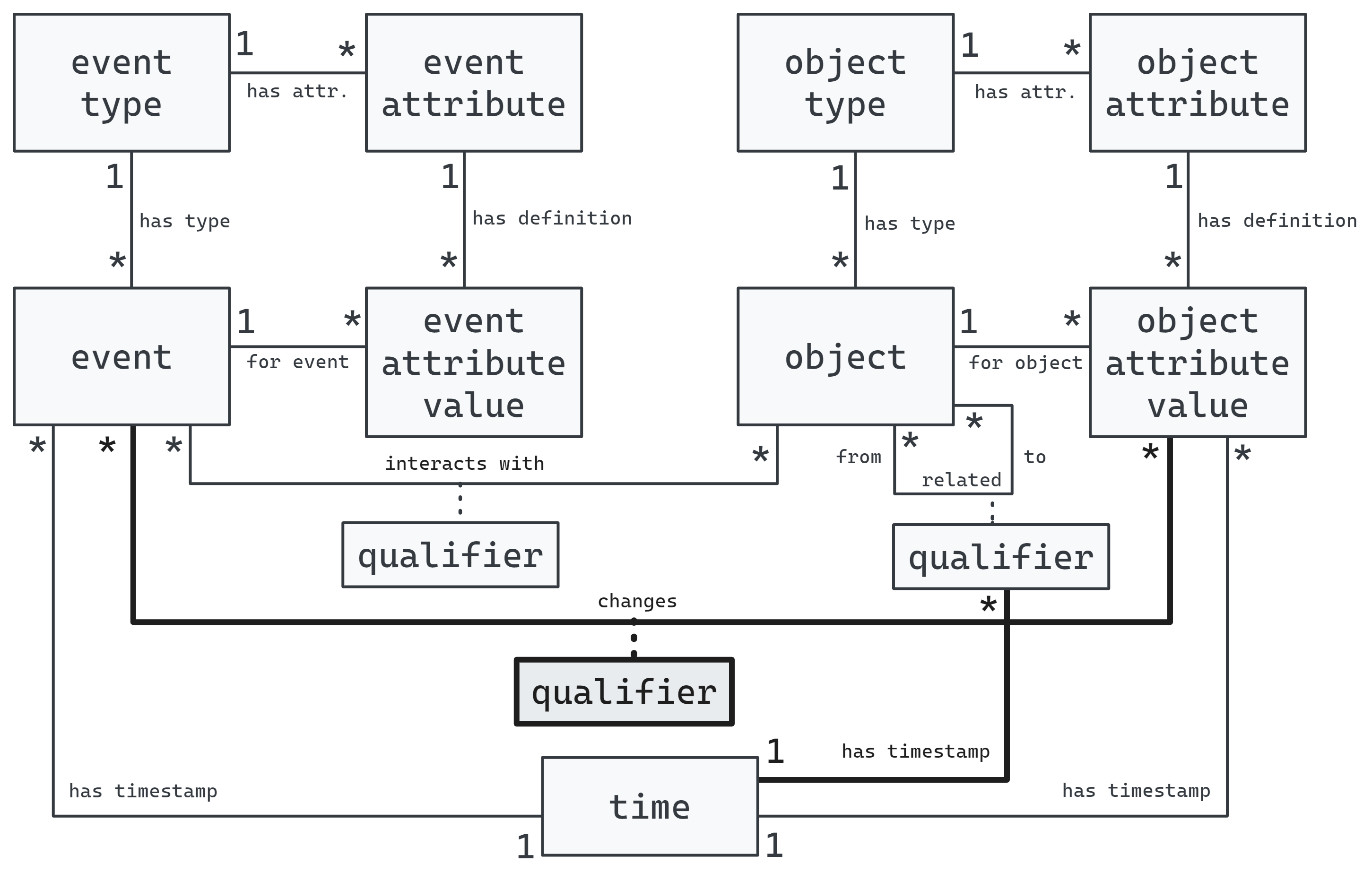}
    \caption{Meta-model of the proposed storage hub.}
    \label{fig:metamodel}
\end{figure}

Figure \ref{fig:metamodel} illustrates our extension of the OCED meta-model used by OCEL 2.0~\cite{vanderAalst_2023,berti2024ocel}.
To fulfill Requirements \ref{Req:E2O} and \ref{Req:O2O}, two additional features are introduced compatible with ACEL and DOCEL, respectively.
First, object-to-object relationships are allowed to change over time by attaching a timestamp to the relationship qualifier. Furthermore, direct relationships between events and object attribute value updates are used to store any known causal relationships between them and thus support many-to-many relationships.  \\

\begin{figure}[!ht]
    \centering
    \includegraphics[width=0.95\textwidth]{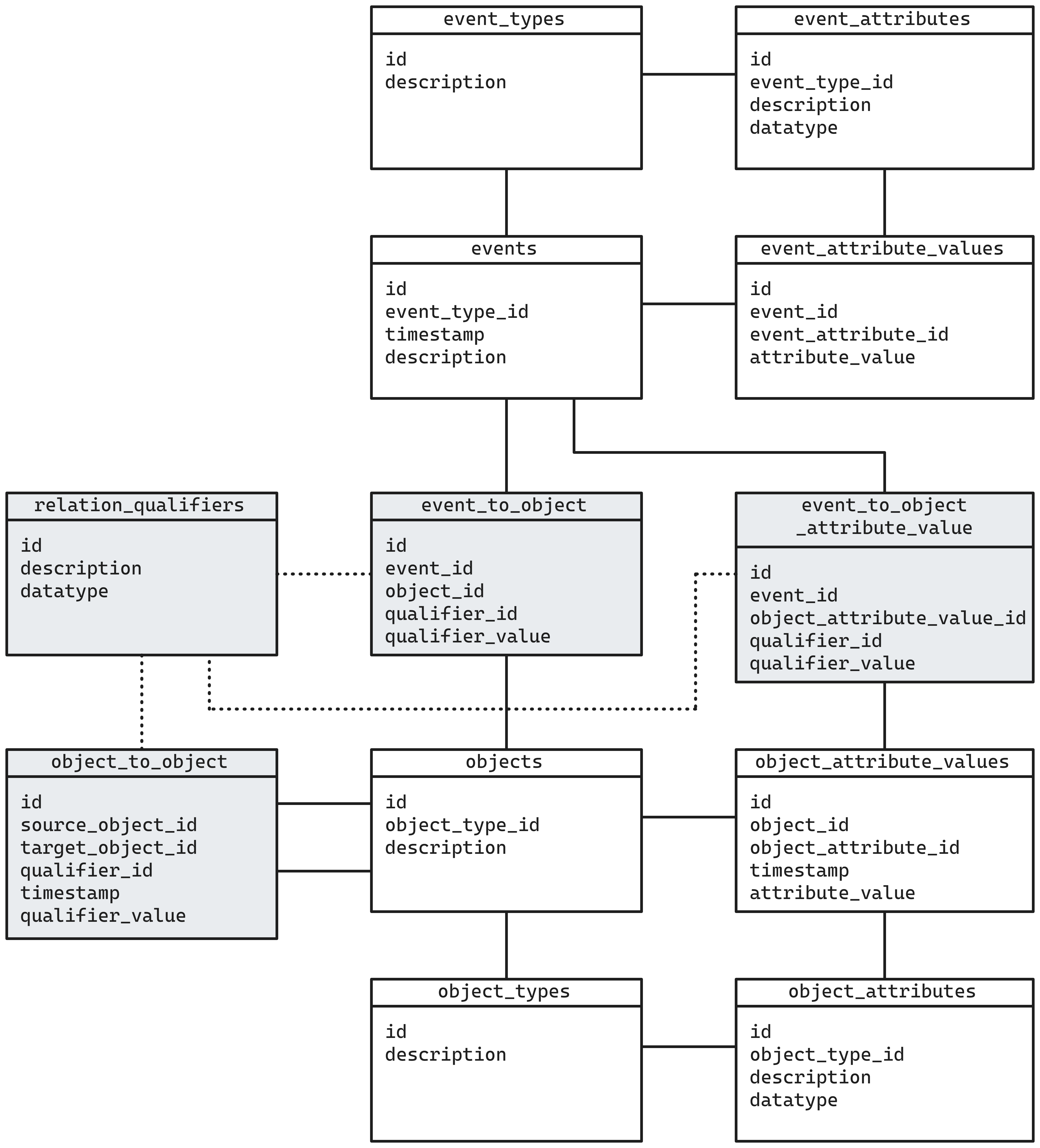}
    \caption{Relational schema for the proposed object-centric event data storage hub, consisting of event tables (top), object tables (bottom) and relation tables (gray).}
    \label{fig:relational_schema}
\end{figure}

Figure \ref{fig:relational_schema} provides an overview of all tables, columns, and relationships employed in the relational schema.
In general, each table incorporates a unique primary key labeled  \texttt{id}.
Foreign keys are constructed by appending `\texttt{\_id}' to the referenced table name. All primary key and foreign key columns are mandatory (non-nullable).
Most tables include a \texttt{description} column to provide human-readable names that can be used for data visualization.
The \texttt{datatype} column is used to keep track of the data type of an attribute or qualifier.
While converting all timestamps to UTC is recommended to mitigate potential time synchronization issues, it is not strictly mandatory.  \\

The relational schema includes four \textbf{event-related tables}, depicted in the upper section of the model in Fig.~\ref{fig:relational_schema}. To maintain flexibility and support dynamic changes (i.e. Requirement \ref{Req:Robust}), event types and their attribute definitions are stored in rows rather than as table and column headers. This approach enables the use of the exact same tables across all processes, reducing the impact of schema modifications.  Changing an event type involves updating foreign keys rather than moving data to different tables, and attributes can be added or removed without altering the schema, in line with Requirement~\ref{Req:Robust}. The four event-related tables are:
\begin{sloppypar}
\begin{itemize}
    \item Table \texttt{event\_types} contains an entry for each unique event type with columns: \texttt{id} and \texttt{description}.
    \item Table \texttt{event\_attributes} stores entries for each unique event attribute with columns \texttt{id}, \texttt{event\_type\_id}, \texttt{description}, and \texttt{datatype}.
    \item Table \texttt{events} records details for each event with columns \texttt{id}, \texttt{event\_type\_id}, \texttt{timestamp}, and \texttt{description}.
    \item Table \texttt{event\_attribute\_values} stores all attribute values for different events with columns \texttt{id}, \texttt{event\_id}, \texttt{event\_attribute\_id}, and \texttt{attribute\_value}.
    This setup decouples events and their attributes by storing each attribute value in a new row, facilitating support for late-arriving data points. This can prove to be an additional advantage when event data originates from multiple systems.
\end{itemize}
\end{sloppypar}

Similar design principles apply to the four \textbf{object-related tables}, depicted in the bottom section of the model in Fig.~\ref{fig:relational_schema}. They also leverage row-based storage to manage attributes independently.  This approach reduces the number of duplicate or NULL values significantly when attributes are updated asynchronously and frequently, in line with Requirement \ref{Req:Scalable}. Due to the nature of IoT data collection\cite{messaging_protocols_for_IOT}, this can be beneficial for industrial use cases.  The four object-related tables are:
\begin{sloppypar}
\begin{itemize} 
    \item Table \texttt{object\_types} records entries for each unique object type with columns \texttt{id} and \texttt{description}.
    \item Table \texttt{object\_attributes} contains entries for each unique object attribute with columns \texttt{id}, \texttt{object\_type\_id}, \texttt{description}, and \texttt{datatype}.
    \item Table \texttt{object} stores details for each object with columns \texttt{id}, \texttt{object\_type\_id}, and \texttt{description}.
    \item Table \texttt{object\_attribute\_values} records attribute values for objects with columns: \texttt{id}, \texttt{object\_id}, \texttt{object\_attribute\_id}, \texttt{timestamp} indicating when attribute updates, and \texttt{attribute\_value} containing the updated value of the attribute.
\end{itemize}
\end{sloppypar}
Finally, the schema includes four \textbf{relation-related tables}. Three of these tables serve as bridging tables to manage the different many-to-many relations between events and objects. The qualifier definitions are stored separately to minimize the impact of renaming them in case of changing business requirements (in line with Requirement \ref{Req:Robust}). It was decided not to include separate tables for qualifiers due to their limited added value.
\begin{sloppypar}
\begin{itemize} 
    \item Table \texttt{relation\_qualifiers} includes three columns: \texttt{id}, \texttt{description}, and \texttt{datatype}.
    In cases where relation qualifiers are not available in the source data, a dummy qualifier can be introduced.
    \item Table \texttt{object\_to\_object} features six columns including \texttt{id}, \texttt{source\_object\_id}, \texttt{target\_object\_id}, \texttt{timestamp} (indicating the validity period of the relationship), \texttt{qualifier\_id} and \texttt{qualifier\_value} (providing additional relationship details). 
    To signify the end of an object-to-object relationship, a NULL value is used for the qualifier value, rather than an end timestamp. This design choice facilitates append-only data ingestion. Only direct relationships between objects are stored, with transitive relations being calculable if necessary (in line with Requirement \ref{Req:Scalable}).
    \item Table \texttt{event\_to\_object} includes columns for  \texttt{id}, \texttt{event\_id}, \texttt{object\_id}, \texttt{qualifier\_id} and \texttt{qualifier\_value}.
    \item Table \texttt{event\_to\_object\_attribute\_value} has five columns for \texttt{id}, \texttt{event\_id}, and \texttt{object\_attribute\_value\_id}, \texttt{qualifier\_id} and \texttt{qualifier\_value}.
\end{itemize}
\end{sloppypar}

Another potential advantage of this setup is that it could simplify anonymization efforts, allowing for relatively easy removal or concealment of type, relation, and attribute descriptions, while working exclusively with anonymous keys.

\section{The \emph{Stack't} Tool}\label{sec:tool}
To demonstrate the practical application of the proposed relational schema supporting a hub-and-spoke architecture for object centric event data within a modern data platform, an open-source tool named \emph{Stack't} was implemented. The source code and documentation for \emph{Stack't} are available on GitHub at \url{github.com/LienBosmans/stack-t}.
\emph{Stack't} is designed to utilize the proposed relational schema as its core data model, aligning with  state-of-the-art data engineering practices that emphasize loose coupling between upstream data sources and downstream applications, robust data lineage, automated testing, and comprehensive documentation.
\emph{Stack't} aims to bridge the gap between industry needs and advancements in object-centric process mining research, offering both interoperability and data engineering best practices to researchers and serving as an entry point for companies interested in adopting  cutting-edge techniques. Below is an overview of the current core capabilities of \emph{Stack't}.
\begin{itemize}
    \item \textbf{Continuous Ingestion:} Supports append-only incremental batch processing of process data.
    \item \textbf{Interactive Visuals:} Provides interactive visualizations for exploratory data exploration
    \item \textbf{Export Capabilities:} Allows exporting to OCEL 2.0, DOCEL, and Neo4j graph database formats.
    \item \textbf{Import from OCEL 2.0:} Includes functionality to import event logs formatted according to OCEL 2.0.
\end{itemize}

\subsection{Data Stack}
\emph{Stack't} leverages a combination of open-source tools within its data stack.
All software components are containerized using Docker, facilitating  virtualization and seamless deployment across laptops, servers or cloud environments.
Data storage and analytics are powered by DuckDB, a high-performance in-process analytical database~\cite{raasveldt2020data}.
Data transformation pipelines are implemented in SQL using dbt-core\footnote{github.com/dbt-labs/dbt-core}, a modern data transformation tool that automates the generation of dbt models from input data.
\emph{Stack't} also generates CSV files suitable for creating graph databases. While Neo4j\footnote{neo4j.com/product/neo4j-graph-database/} is not integrated directly into \emph{Stack't}, it can be deployed in a separate Docker container.


\subsection{Data Import-Export}
\emph{Stack't} provides Python scripts for importing and exporting event logs in various formats. It supports importing and exporting  OCEL 2.0 event logs~\cite{berti2024ocel} as SQLite databases and exporting event logs in DOCEL~\cite{goossens2022enhancing} as CSV files. 
Additionally, for backward compatibility, \emph{Stack't} includes an option to export event logs formatted for a single object type as CSV files.
DuckDB's native support for a wide range of data formats—including CSV, Parquet, JSON, Excel, MySQL, PostgreSQL, and SQLite—ensures seamless data ingestion from diverse sources\footnote{DuckDB documentation for recommended data import methods for each supported format can be found at duckdb.org/docs/guides/import/overview.}.
Ensuring traceability and maintaining the provenance of process data~\cite{TerHofstede2023, Martin2021} is integral to the design of \emph{Stack't}. Leveraging dbt's built-in data lineage capabilities, \emph{Stack't} optimizes traceability by tracking the origin and transformations applied to data. 
DuckDB, in conjunction with dbt, facilitates the data engineering processes necessary to extract process data from heterogeneous sources and load it into our relational schema.
As an example, \ref{sec:jaffle-shop} includes a comprehensive source-to-target mapping for a synthetic relational database simulating customer behavior in jaffle shops\footnote{The dataset was obtained using \texttt{jafgen}, an open-source tool for generating synthetic data for analytics engineering practice and demonstration. Source code available on github.com/dbt-labs/jaffle-shop-generator.}, which is also available on the Stack’t repository .
During our development process, two key insights were encountered. First, translating OCED meta-models into code posed challenges due to the lack of standardized naming conventions, which currently limits support for importing OCED logs to OCEL 2.0. Second, direct copying of data from source systems into a storage hub is often impractical, as most application databases are not optimized for analytical use cases. This necessitates additional efforts, such as calculating relevant attributes, to align data from its system representation with real-life process representations.

\subsection{Automated Data Quality Tests}
Automated tests serve as critical checkpoints throughout our data pipeline, ensuring robustness and consistency. These tests are integrated with dbt models documented in YAML files.
Following ingestion (staging) of an OCEL 2.0 event log, the initial checkpoint verifies the following conditions:
\begin{enumerate}
    \item \textbf{Unique Primary Keys:} Ensures that primary keys are unique and not null.
    \item \textbf{Foreign Key Constraints:} Validates that foreign keys are not null.
    \item \textbf{Referential Integrity:} Confirms foreign keys exist as primary keys in their referenced tables.\label{item:test_missing_keys}
    \item \textbf{Timestamp Validity:} Verifies that timestamps are not null.
\end{enumerate}

\begin{sloppypar}
These tests are detailed in the \textit{`staging\_models.yml'} file located in the \textit{`dbt\_project/models'} folder of our code repository.
The testing process is validated using three published OCEL 2.0 event logs in SQLite format: \texttt{Container Logistics} \cite{knopp_2023_8428084}, \texttt{Order Management} \cite{knopp_2023_8428112}, and \texttt{Procure-To-Payment} \cite{park_2023_8412920}.
For two out of three datasets, the test for referential integrity (\ref{item:test_missing_keys}) failed because not all referenced objects exist in the \texttt{object} table.
Missing objects in \texttt{Container Logistics} and \texttt{Procure-To-Payment} were identified, and subsequently added to our data pipeline before mapping the data to the hub.
Additionally, an issue with SQLite not enforcing column types was encountered, where integer columns contained 'null' values of type text instead of standard NULL values\footnote{This bug was reported and has been resolved by the creators of Ocelot with version release 0.1.2 - 2024-01-16 (ocelot.pm/about).}.
Both solutions are documented in detail in the \textit{`README'} file of our code repository.  \\
\end{sloppypar} 

A second checkpoint is placed after all process data is mapped to our relational schema, reaffirming the integrity of our data transformations.
The \textit{`transform\_models.yml'} file verifies points 1-4 once more to ensure that the data mapping was done correctly.
Since these tests run automatically whenever the data pipeline is used to transform data, they continuously enforce the relational schema of the hub, avoiding downstream data quality issues caused by inconsistencies.

The final checkpoint precedes the generation of input files for a graph database, as described in the file \textit{`graph\_models.yml'}.
These tests validate the compatibility of generated CSV files with Neo4j, ensuring all nodes have unique identifiers and all edges link to existing nodes.

\subsection{Application: Interactive Graph Visualizations}
\begin{figure}[!htbp]
    \centering
    \includegraphics[width=0.90\textwidth]{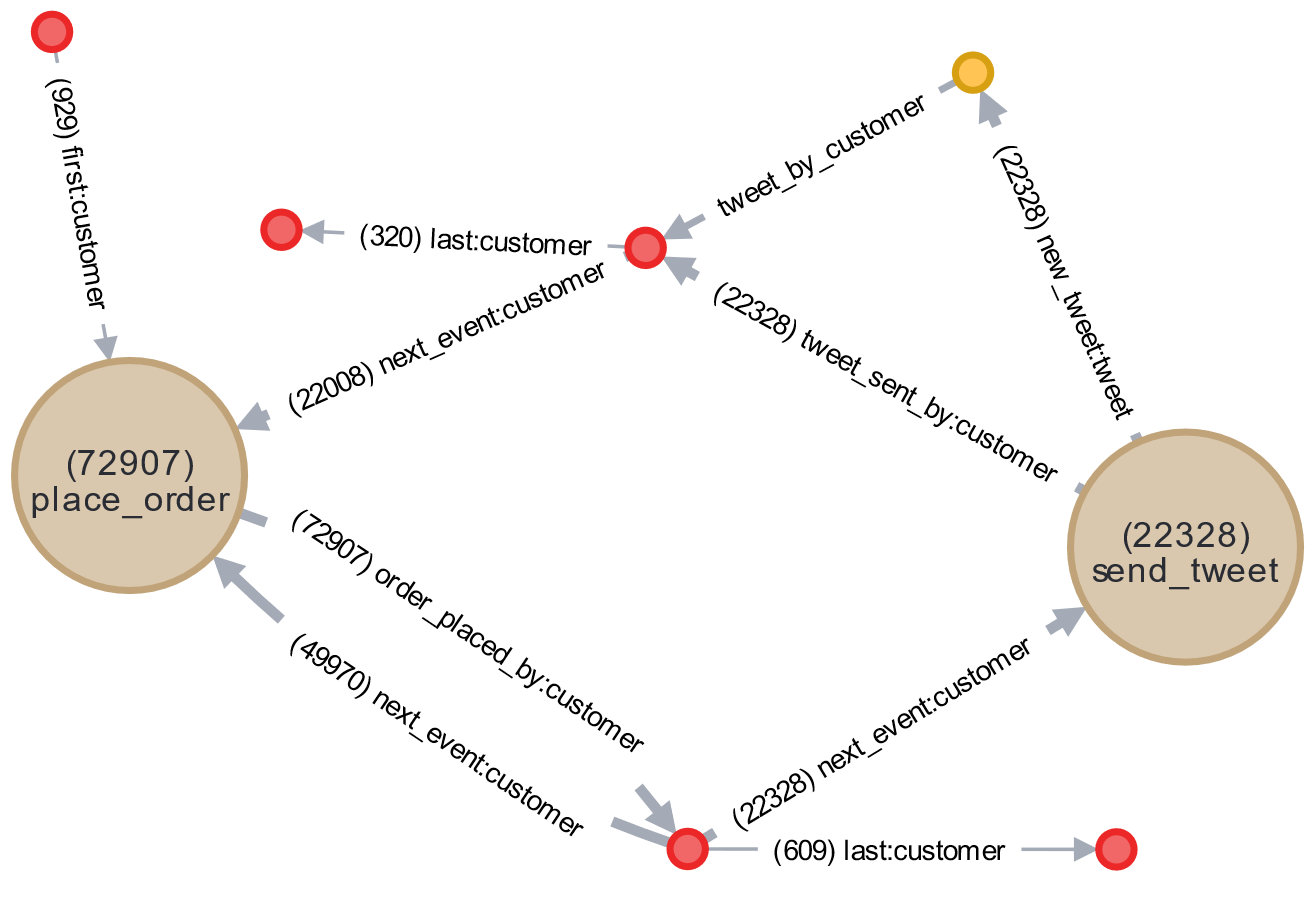}
    \caption{Example of interactive process data visualizations in Neo4j.}
    \label{fig:screenshot_stackt}
\end{figure}

Once all process data is mapped to the hub, \emph{Stack't} facilitates multiple applications for data exploration and visualization. In the current implementation, the tool provides data pipelines for generating two types of information.
Firstly, \emph{Stack't} can produce overview tables summarizing the process data. 
These tables contain an overview of how many events (objects) exist for each event (object) type, and statistics on the different relationships.
Additionally, \emph{Stack't} offers a data pipeline that outputs CSV files capable of generating a graph database in Neo4j.
These files define a prototype implementation for visualizing processes captured through object-centric event data. An example using synthetic customer data is illustrated in Figure~\ref{fig:screenshot_stackt}.
It is important to note that this visualization approach differs from traditional process model visualizations found in literature, such as colored Petri nets or object-centric Petri nets~\cite{Jensen2007,Adams2022}.

The visualizations are designed for interactive data exploration, supporting process owners to validate whether the data mapping has been done appropriately. Because the graph is accessed through Neo4j, filtering and interaction can be done through the Cypher query language.
In the current setup, object-to-object relations are established only between object snapshots that have the same timestamp, which may present a limitation.
Individual case visualizations can be enhanced by grouping object snapshots based on the combination of their object type, the event type of the previous event, and the set of updated attributes. 
More details on the visualization setup are described in \ref{sec:vis_tool}.
The exporters provided by \emph{Stack't} enable process owners to obtain event logs compatible with different formats, facilitating the use of existing process discovery algorithms to derive (more classical) process models, of any process language.

\section{Conclusion}\label{sec:conclusion}
In this paper, motivation is provided for the introduction of a hub-and-spoke data ingestion and storage architecture tailored for organizations adopting object-centric process mining. By decoupling data sources from applications, this architecture offers flexibility and maintainability, facilitating continuous ingestion of new process data and integration of new data types, without compatibility issues, and requiring setup only once.
Therefore a relational schema is introduced, designed specifically for such an intermediate data storage hub, guided by practical considerations such as scalability and maintainability.  \\

Moreover, our flexible relational schema also accommodates temporal and scalable (thanks to only storing direct) object-to-object relationships, features not commonly supported by existing object-centric event log storage formats, while ensuring compatibility with leading formats such as OCEL 2.0.
This design prioritizes modularity and generalizability, enabling extraction of data in different formats and granularities as needed by specific applications. 
Thus, this work serves as a blueprint for constructing robust and maintainable data pipelines in the context of object-centric process mining.
To materialize our approach, \emph{Stack't} was developed, an open-source tool that implements our proposed relational schema within a modern data stack. \emph{Stack't} serves as an accessible entry point for organizations looking to explore object-centric process mining, further bridging the gap between industry applications and process mining research.  \\

In addition, it is important to note that our current relational schema primarily supports structured data. Future research could explore methods to label and integrate unstructured data sources, such as e-mails or videos, into our database format, which could in turn enable development of novel process mining techniques that can combine unstructured and structured data sources. Additionally, enhancing our tool’s data visualization capabilities to display concurrent events more intuitively and expand support for object-to-object relations across different timestamps represents avenues for future development. Furthermore, improving data import and export functionalities to seamlessly integrate with diverse data sources and OCED formats would enhance the tool's utility and interoperability.  \\

In conclusion, our work lays a foundation for advancing object-centric process mining capabilities through an innovative hub-and-spoke architecture and corresponding open-source tooling. We believe this framework can catalyze fruitful collaborations between academia and industry. We invite researchers and practitioners to join us in advancing and refining these initiatives together.

\bibliographystyle{elsarticle-num} 
\bibliography{ref.bib}


\newpage
\appendix

\section{Source-to-target data mapping example for a running example}
\label{sec:jaffle-shop}

This appendix describes how \textit{Stack't} supports data preparation for object-centric process mining, using a running example of a synthetic dataset in relational database format. 
From this database, 95,241 events, 96,209 objects, 263,383 event-to-object relations, 343,343 object-to-object relations, and 929 event-to-object-attribute-value relations are extracted.
All functionality described in this appendeix is included in the repository as a template which works out-of-the-box. On a laptop it typically takes less than 3 minutes to ingest the source data, map it to the relational schema and export it to an OCEL 2.0 event log.\footnote{Depening on the hardware. A Windows laptop with Intel(R) Core(TM) i7-8565U CPU and 16GB of RAM, using Windows Subsystem for Linux (WSL) to run the Docker container, needs less than 2 minutes.}

\subsection{Generating source data}

The synthetic datasetis generated using \textit{jaffle shop generator}, an open-source project hosted on GitHub by dbt-labs\footnote{https://github.com/dbt-labs/jaffle-shop-generator}. A description of the data, copied from the repository's README file, can be found below.

\begin{quote}
    "The Jaffle Shop Generator or \texttt{jafgen} is a simple command line tool for generating synthetic datasets suitable for analytics engineering practice or demonstrations. The data is generated in CSV format and is designed to be used with a relational database. It follows a simple schema, with tables for:
    \begin{itemize}
        \item Customers (who place Orders)
        \item Orders (from those Customers)
        \item Products (the food and beverages the Orders contain)
        \item Order Items (of those Products)
        \item Supplies (needed for making those Products)
        \item Stores (where the Orders are placed and fulfilled)
        \item Tweets (Customers sometimes issue Tweets after placing an Order)
    \end{itemize}
    
    It uses some straightforward math to create seasonality and trends in the data, for instance weekends being less busy than weekdays, customers having certain preferences, and new store locations opening over time. We plan to add more data types and complexity as the codebase evolves."
\end{quote}

The \textit{Stack't} repository includes a generated dataset of one year simulated data, stored in the folder \textit{`stack-t/event\_data/jaffle\_shop'}. Instructions to create a different or larger dataset are included in the README file located inside the same folder.

\subsection{Connecting to source data}

The connection to the source tables is described in a YAML file of which an excerpt for the first table is shown in code block \ref{code:sources.yml}. The \texttt{external\_location} option is used to define the connection to the correct CSV file. Note that this option can also be used to connect to files hosted online, e.g. a parquet file stored in a data lake.

\begin{lstlisting}[breaklines,caption={Excerpt from \textit{`stack-t/dbt\_project/models/sources.yml'}},label={code:sources.yml}]
sources:
  - name: jaffle_shop
    tables:
      - name: customers
        description: Customers (who place Orders)
        meta:
          external_location: "read_csv('../event_data/jaffle_shop/jaffle_customers.csv',delim=',',header=true,auto_detect=true)"
\end{lstlisting}

\subsection{Ingesting source data}

To make full usage of the analytical capabilities of DuckDB, the source data is loaded into a DuckDB database file first. This step is commonly referred to as \textit{staging} the data. The database file \textit{`dev.duckdb'} is stored inside the folder \textit{`stack-t/dbt\_project'} and the connection to the database file is defined in a YAML file \textit{`profiles.yml'} that can be found in the same folder.

Since the CSV files are ingested as data tables without any modifications, the SQL code is very straightforward as can be observed in code block \ref{code:stg_customers.sql} for an example table. The SQL code used for staging can be found inside the folder \textit{`stack-t/dbt\_project/models/staging'}.

\begin{lstlisting}[breaklines,caption={\texttt{stack-t/dbt\_project/models/staging/stg\_customers.sql}},label={code:stg_customers.sql}]
select * from {{ source('jaffle_shop','customers') }}
\end{lstlisting}

\subsection{Source-to-target mapping}

The source-to-target mapping was done in two stages. First, all the different event and object types, their attributes, and the type of relationships between them were identified using using a five-step approach which is described after this paragraph. Afterwards, the SQL code of the source-to-target mapping is written and saved inside the folder \textit{`stack-t/dbt\_project/models/transform'}. The code itself is not included in this appendix but documentation is available online.

\paragraph{Step 1: Identifying events} Using timestamp columns in the source data as a guide, three event types were identified: \textit{place\_order}, \textit{open\_store} and \textit{send\_tweet}, found respectively in tables \texttt{orders}, \texttt{stores} and \texttt{tweets}.

\paragraph{Step 2: Linking events to objects} Exploring primary and foreign keys in the database led to the identification of the following object types and event-to-object relationships.
\begin{itemize}
    \item Event type \textit{place\_order} is related to object types \textit{order}, \textit{store} and \textit{customer}.
    \item Event type \textit{open\_store} is related to the object type \textit{store}.
    \item Event type \textit{send\_tweet} is related to object types \textit{customer} and \textit{tweet}.
\end{itemize}

\paragraph{Step 3: Linking objects to objects} Exploring the foreign key columns in the tables describing object types, the following object-to-object relationships were identified.
\begin{itemize}
    \item Object type \textit{order} is related to \textit{product}, \textit{customer} and \textit{store}.
    \item Object type \textit{product} is related to \textit{ingredient}.
    \item Object type \textit{customer} is related to \textit{tweet}.
\end{itemize}

\paragraph{Step 4: Identifying attributes} For each event and object type, attributes of interest are identified. Note that these don't need to be available directly in the source data but can be calculated instead. Below is an overview of all attributes. An asterisk * is used to indicate attributes that are not available directly in the data and therefore calculated based on other data available.
\begin{itemize}
    \item Event type \textit{place\_order} has attribute \textit{total\_is\_correct}*.
    \item Object type \textit{order} has attributes \textit{subtotal}, \textit{tax}, \textit{total} and \textit{item\_count}*.
    \item Object type \textit{product} has attributes \textit{price}, \textit{type}, \textit{description}, \textit{cost}* and \textit{margin\_perc}*.
    \item Object type \textit{tweet} has attribute \textit{content}.
    \item Object type \textit{store} has attributes \textit{tax\_rate} and \textit{customer\_count}*.
    \item Object type \textit{ingredient} has attributes \textit{cost} and \textit{is\_persishable}.
    \item Object type \textit{customer} has attributes \textit{tweet\_count}* and \textit{favorite\_product}*.
\end{itemize}
After discovering that a customers favorite product is an important attribute, it was decided to replace this object attribute with a dynamic object-to-object relationship between object types \textit{customer} and \textit{product} instead. 

\paragraph{Step 5: Linking events to object attribute value updates} After defining the different attributes, relationships between events and the change of an object attribute value can be identified.
\begin{itemize}
    \item Event type \textit{send\_tweet} directly affects the attribute \textit{tweet\_count} of object type \textit{customer}.
    \item Event type \textit{place\_order} affects the attribute \textit{customer\_count} of object type \textit{store} when it is the first order placed by the customer at this store.
\end{itemize}

\newpage
\paragraph{Summary} Here a comprehensive overview of all identified event and object types, their attributes, and the type of relationships between
them is given.
\begin{itemize}
    \item Event types (with event attributes):
    \begin{itemize}
        \item \textit{place\_order} (\textit{total\_is\_correct})
        \item \textit{open\_store}
        \item \textit{send\_tweet}
    \end{itemize}
    \item Object types (with object attributes):
    \begin{itemize}
        \item \textit{order}(\textit{subtotal}, \textit{tax}, \textit{total}, \textit{item\_count})
        \item \textit{product} (\textit{price}, \textit{type}, \textit{description}, \textit{cost}, \textit{margin\_perc})
        \item\textit{tweet} (\textit{content})
        \item \textit{store} (\textit{tax\_rate}, \textit{customer\_count})
        \item \textit{ingredient} (\textit{cost}, \textit{is\_persishable})
        \item \textit{customer} (\textit{tweet\_count})
    \end{itemize}
    \item Event-to-object relationship types (from event type - to object type):
    \begin{itemize}
        \item \textit{new\_order} (\textit{place\_order} - \textit{order})
        \item \textit{ordered\_product} (\textit{place\_order} - \textit{product})
        \item \textit{order\_placed\_in} (\textit{place\_order} - \textit{store})
        \item \textit{order\_placed\_by} (\textit{place\_order} - \textit{customer})
        \item \textit{new\_store} (\textit{open\_store} - \textit{store})
        \item \textit{new\_tweet} (\textit{send\_tweet} - \textit{tweet})
        \item \textit{tweet\_sent\_by} (\textit{send\_tweet} - \textit{customer})
    \end{itemize}
    \item Object-to-object relationship types (from object type - to object type)
    \begin{itemize}
        \item \textit{order\_placed\_in\_store} (\textit{order} - \textit{store})
        \item \textit{order\_placed\_by\_customer} (\textit{order} - \textit{customer})
        \item \textit{order\_contains\_product} (\textit{order} - \textit{product})
        \item \textit{ingredient\_used\_for\_product} (\textit{ingredient} - \textit{product})
        \item \textit{tweet\_by\_customer} (\textit{tweet} - \textit{customer})
        \item \textit{customer\_favorite\_product} (\textit{customer} - \textit{product})
        \item \textit{customer\_of\_store} (\textit{customer} - \textit{store})
    \end{itemize}
    \item Event-to-object-attribute-values relationship types (from object type - to object attribute, of object type)
    \begin{itemize}
        \item \textit{another\_tweet} (\textit{send\_tweet} - \textit{tweet\_count}, \textit{customer})
        \item \textit{first\_store\_visit} (\textit{place\_order} - \textit{customer\_count}, \textit{store})
    \end{itemize}
\end{itemize}

Note that the flexibility of the relational schema allows to expand and refine the chosen types, attributes and relations with ease, encouraging an iterative approach. Nevertheless, we believe these steps can provide a good starting point for data exploration and preparation.

\subsection{Extractor for OCEL 2.0}

While the relational schema of OCEL 2.0 is clearly defined in \cite{berti2024ocel}, a number of tables are dependent on the event and object types, and their attributes. Therefore implementing the extractor as straight-forward SQL code was not possible.
This was resolved by using a Python script that automatically generates and runs the SQL code needed for extracting the data as an OCEL 2.0 log  The resulting SQLite database file is written to the folder \textit{`stack-t/exports/ocel2'}.

Since the meta-model used by \textit{Stack't} is more flexible then what is allowed by OCEL 2.0, the following design choices were made for the data mapping.
\begin{itemize}
    \item All event-to-object-attribute-value relations are ignored.
    \item All object-to-object relations are assumed to be static. The description of the object-to-object relation qualifier is used instead of the qualifier value.
\end{itemize}

A high-level overview of the Python script used to extract OCEL 2.0 logs:
\begin{itemize}
    \item DuckDB (as in-memory database, using the Python Client API) is used to read data from the \textit{Stack't} relation schema stored in the DuckDB database file, to transform it into an OCEL 2.0 relational schema, and to write the result to a SQLite database file. To this end, both a DuckDB and an empty SQLite database are attached to the in-memory database.
    \item The tables that are not dependent on the specific process, i.e. all tables except \texttt{object\_<type>} and \texttt{event\_<type>} tables, are pre-calculated by the data pipeline\footnote{The relevant SQL code is stored in \textit{`stack-t/dbt\_project/models/export/ocel2'}.}. Therefore these tables can be copied directly to the SQLite database.
    \item A double for-loop is used to create the SQL code (as a multiline string) for the different \texttt{event\_<type>} tables. The first for-statement loops over the list of all event types, which are needed to create the tables. The second for-statement loops over the event attributes of each event type in order to create the table columns. The unpivoted data is pre-calculated in the data pipeline. Therefore it only needs to be filtered and pivoted before being copied to each of the newly created tables in SQLite.
    \item The same approach is used for the \texttt{object\_<type>} tables.
\end{itemize}

All code, including documentation, can be found online in the repository. The Python script \textit{`export\_to\_ocel2.py'} is stored in folder \textit{`stack-t/python\_code'}. Other extractors, to DOCEL and (non-object-centric) CSV files, were implemented using a similar appraoch and can be found in the same folder

\newpage

\section{Visualization Tool}
\label{sec:vis_tool}

As described in the main body of this work, the tool allows exporting the process data to a graph database in Neo4j.
By following certain rules, described below, the process data can be visualized and consequently be used for interactive data exploration, with the purpose of e.g. checking the data mapping.
Instead of a classic directly-follow event graph that assumes that every event is linked to exactly one object, multiple objects are supported and therefore also multiple directly-follow edges can start from a single event.
To differentiate between the different objects, object snapshot nodes are added, next to event nodes.
An object snapshot represents an object at a certain timestamp.
Additional object snapshot nodes are added when object attribute updates don't share a timestamp with a related event.
Finally, object-to-object edges are added between object snapshots with the same timestamp for which an object-to-object relationship is valid at that moment.
\begin{figure}[htbp]
    \centering
     \includegraphics[width=0.99\textwidth]{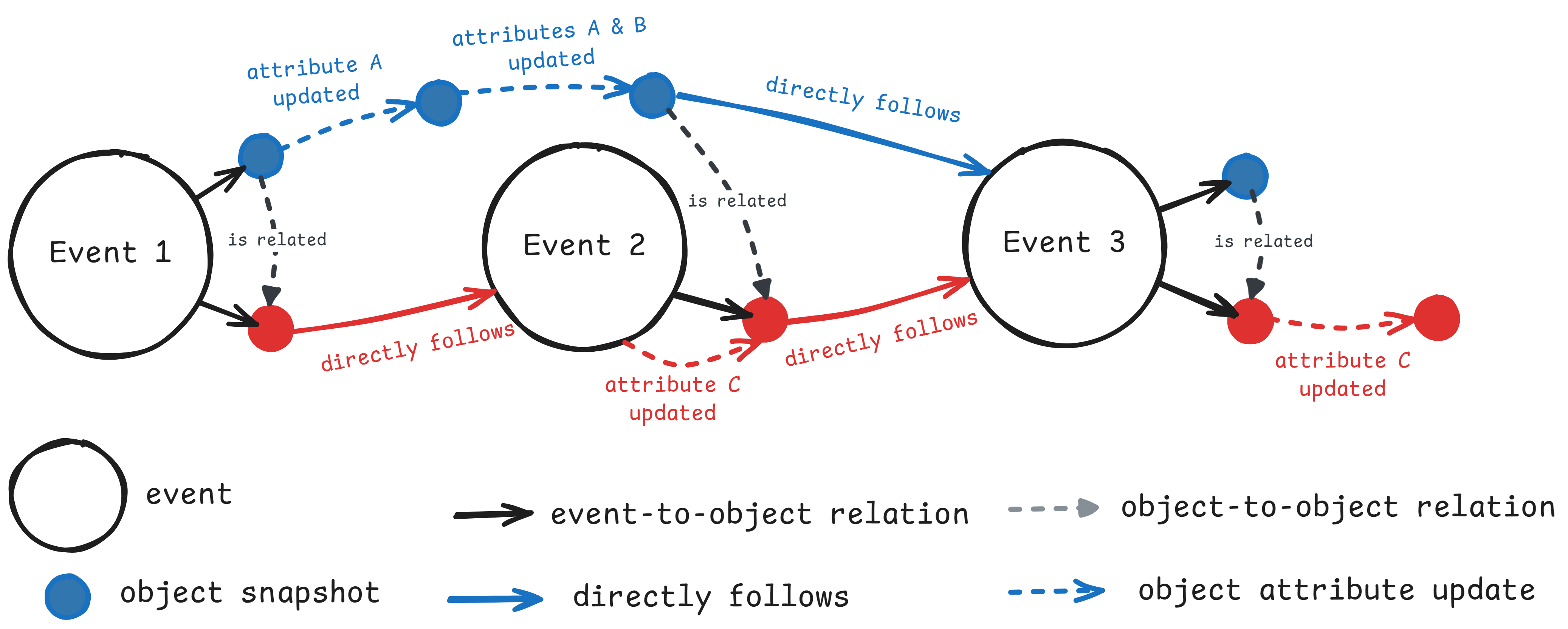}
     \caption{Current prototype of the process data visualization.}
     \label{fig:legend_individual_visualizations}
 \end{figure}

 \begin{figure}[htbp]
     \centering
     \includegraphics[width=0.93\textwidth]{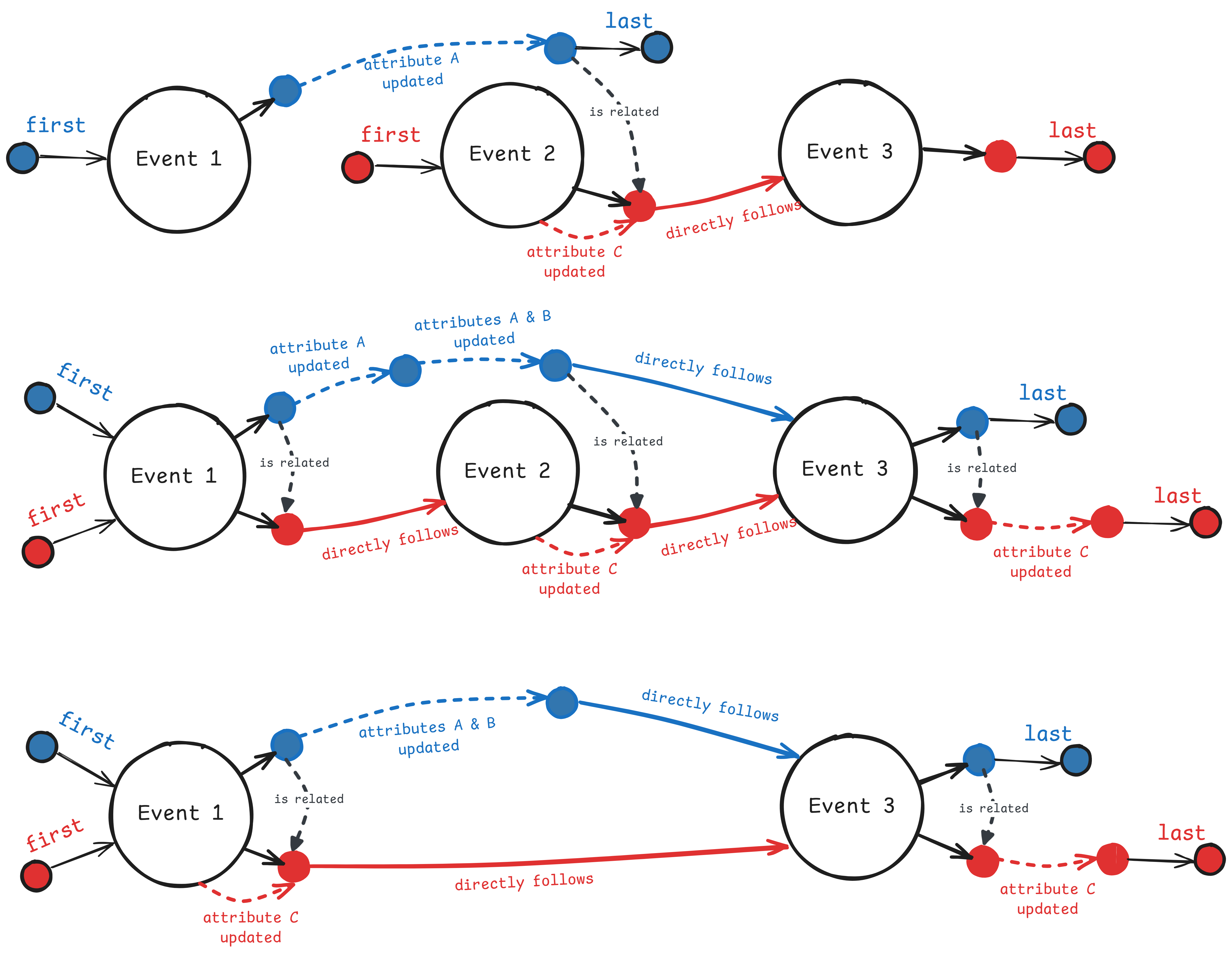}
     \caption{Three examples of a case within a process that uses three event types (1,2,3), a blue object with two attributes (A,B) and a red object with one attribute (C).}
     \label{fig:example_individual_visualizations}
 \end{figure}

\begin{figure}[htbp]
     \centering
     \includegraphics[width=0.93\textwidth]{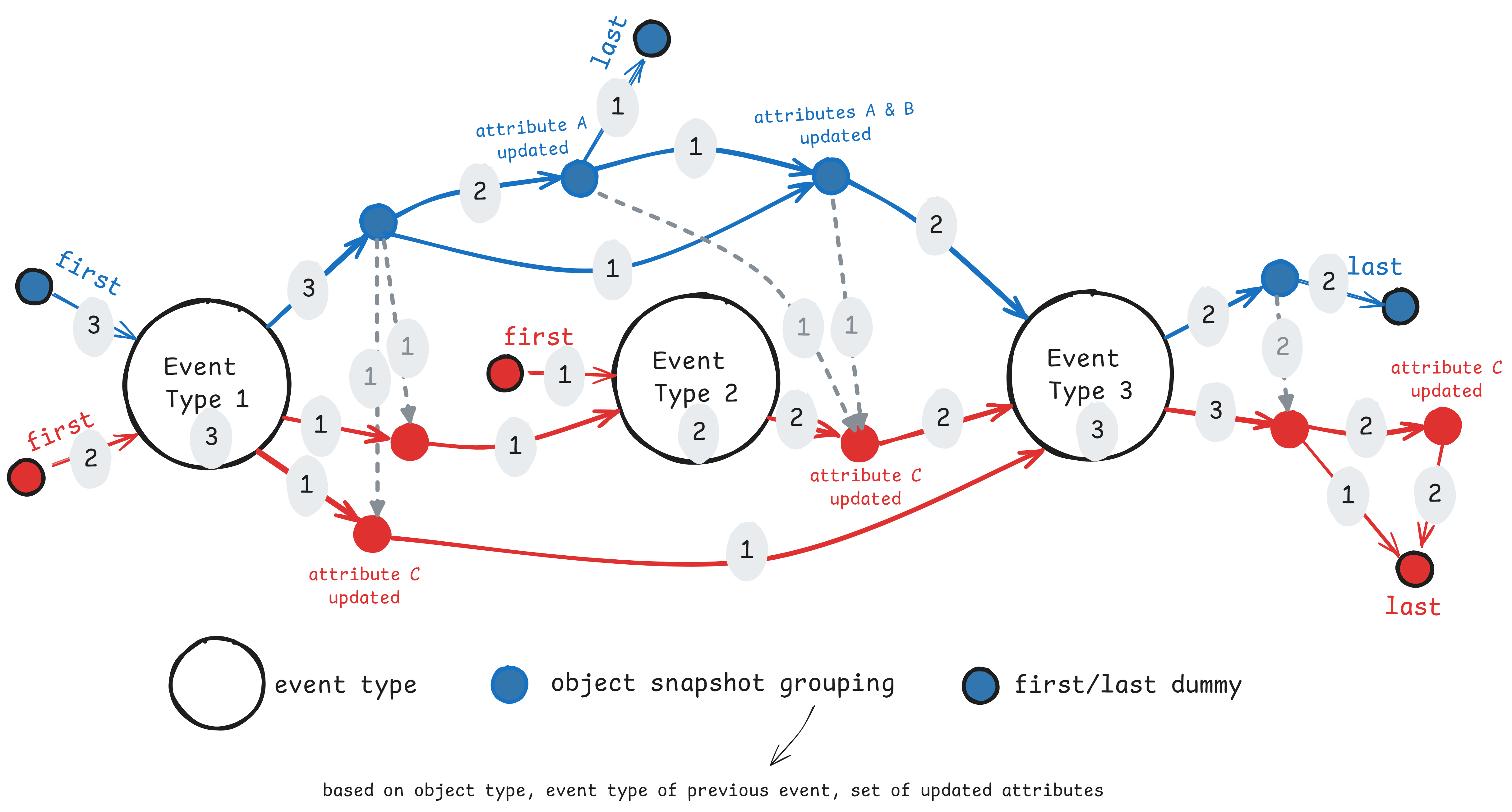}
     \caption{Overview visualization of the three cases shown in figure \ref{fig:example_individual_visualizations}.}
     \label{fig:example_overview_visualizations}
\end{figure}

Concurrency is (currently) not supported. Instead, concurrent events are linked one after the other, ordered by \texttt{event\_type\_id} (first) and \texttt{event\_id} (second).
Note that object-to-object relations are only added between object snapshots that have the same timestamp, which might be a limitation.
Figure~\ref{fig:example_individual_visualizations} shows an example of three selected cases within a process. To show summary process visualization, over multiple cases, nodes are mapped to \textit{overview nodes}; events are mapped to their event types and object snapshots are mapped to \textit{object snapshot groupings}. 
For this, multiple strategies are possible. In the current setup, it was opted to group object snapshots based on the combination of their object type, the event type of the previous event, and the set of updated attributes.
Edges are mapped to \textit{overview edges} based on the mapping of their start and end node. 
Figure~\ref{fig:example_overview_visualizations} shows how the cases from Figure~\ref{fig:example_individual_visualizations} are combined into a single process overview visualization.
For a more comprehensive understanding of this functionality, including detailed examples, please refer to our online resources accessible via \url{github.com/LienBosmans/stack-t}.

\end{document}